# Deep-Learning Discovers Macroscopic Governing Equations for Viscous Gravity Currents from Microscopic Simulation Data

**Junsheng Zeng[1], Hao Xu[2], Yuntian Chen[1], and Dongxiao Zhang[3*]**

[1] Frontier Research Center, Peng Cheng Laboratory, Shenzhen 518000, P. R. China

[2] College of Engineering, Peking University, Beijing 100871, P. R. China

[3] School of Environmental Science and Engineering, Southern University of Science and Technology, Shenzhen 518055, P. R. China

*Corresponding author: Dongxiao Zhang (zhangdx@sustech.edu.cn)

**Key points**

- Macroscopic governing equation is discovered on the basis of high-resolution microscopic simulation data without the need of specifying the potential terms.

- Deep-learning framework captures the dominated terms which agree with theoretically derived equations.

- Compensation PDE terms are discovered for interpreting the deviation between theoretical predictions and simulation results.

# Abstract

Although deep-learning has been successfully applied in a variety of science and engineering problems owing to its strong high-dimensional nonlinear mapping capability, it is of limited use in scientific knowledge discovery. In this work, we propose a deep-learning based framework to discover the macroscopic governing equation of viscous gravity current based on high-resolution microscopic simulation data without the need for prior knowledge of underlying terms. For two typical scenarios with different viscosity ratios, the deep-learning based equations exactly capture the same dominated terms as the theoretically derived equations for describing long-term asymptotic behaviors, which validates the proposed framework. Unknown macroscopic equations are then obtained for describing short-term behaviors, and additional deep-learned compensation terms are eventually discovered. Comparison of posterior tests shows that the deep-learning based PDEs actually perform better than the theoretically derived PDEs in predicting evolving viscous gravity currents for both long-term and short-term regimes. Moreover, the proposed framework is proven to be very robust against non-biased data noise for training, which is up to 20%. Consequently, the presented deep-learning framework shows considerable potential for discovering unrevealed intrinsic laws in scientific semantic space from raw experimental or simulation results in data space.

# 1 Introduction

Deep-learning algorithms are currently being successfully applied in numerous science and engineering fields, such as physics[1,2], earth science[3,4], and computer vision and speech recognition[5]. The advantage of deep-learning algorithms is their strong capability of constructing high-dimensional nonlinear mappings. Therefore, a deep-learning based surrogate model may be trained to effectively describe a physical process based on collected observation data. However, this type of surrogate model constitutes a "black box" with poor explainability, and fails to deepen our understanding of the essences of the physical process, which greatly decreases its utility.

This constitutes a challenging problem, which combines knowledge discovery and explainable machine learning[6]. Humans' understanding of the world is based on semantic space. In semantic space, people are able to understand and construct rules for accurate inferences and analyses. In contrast, pure data-driven surrogates exist in data space. In data space, the expressive capability of surrogates is strong with infinite possibilities. It is particularly challenging, however, to interpret useful information hidden in amounts of data. Therefore, the key to explainability for a machine-learning based surrogate is to transform high-dimensional mapping from data space into semantic space, and obtain an intuitive and interpretable understanding about the mapping. In modern sciences, the most widely acceptable semantic form or knowledge expression form is represented by partial/ordinary differential equations (PDEs/ODEs), which are usually concise and easy-to-understand.

To realize the transformation of surrogates from data space into semantic space, data-driven discovery of partial differential equations (PDEs) is a feasible solution, and thus has recently attracted increasing attention. Usually, PDEs consist of several complex differential terms, and the candidate library of potential terms can be very large. Determination of how to accurately reproduce the differential terms and find parsimonious PDEs become key objectives. Recently, Xu et al.[7] proposed a deep learning-genetic algorithm (DLGA) framework. In this framework, a deep neural network is utilized to calculate derivatives and generate meta-data, and the genetic algorithm is employed to discover the form of PDEs without the need of including the true terms in the initial guess. The deep neural network has been utilized in PDE discovery in previous works[8–10], because the derivatives calculated by automatic differentiation are more accurate and robust to noise. In the process of the genetic algorithm, genomes are composed of several basic genes, which can be adjusted according to the situation of the discovery process, which markedly increases both flexibility and practicability. Compared to other sparse regression methods, including LASSO[11], SINDY[12,13], sequential threshold ridge regression (STRidge)[14,15], and sparse Bayesian regression[16], the genetic algorithm does not need a complete candidate library beforehand[17], which may be impossible for many real-world applications. Meanwhile, the integral form is an efficient approach to facilitate the PDE discovery process and increase the accuracy of discovered PDEs[18].

In this work, we attempt to employ the deep-learning based PDE discovery framework[18] to discover unknown macroscopic equations for a real physical process, i.e., viscous gravity current. Viscous gravity current is an important natural phenomenon in geophysics[19–21], and typical scenarios include displacement flows in oil reservoirs[22], sea water intrusion[23,24], water injection in geothermal reservoirs[25], pollutant dispersion through groundwater[26], and proppant transport in hydraulic fractures[21,27]. Essentially, viscous gravity current is a gravity-driven flow constrained in porous media or narrow vertical fractures due to the density difference between intruding and *in-situ* fluid, which forms a remarkable sharp interface at macro-scale, i.e., the current front. As a consequence, identification of a macroscopic equation for describing the evolution of current front height is a crucial issue. In extant literature, there exist many theoretically derived macroscopic equations for viscous gravity currents under various conditions, such as different viscosity ratio[19], multi-layer porous media[28], non-Newtonian intruding fluid[29–31], and varying horizontal permeability[32]. However, these works largely focus on long-term asymptotic behaviors of viscous gravity currents, and macroscopic equations for capturing short-term behaviors remain undetermined.

Different from previous works in data-driven PDE discovery, the training data of current front height are directly extracted from high-resolution microscopic simulation results instead of solving already-known model equations, and the long-term theoretical PDEs are only utilized in posterior tests as a reference. Consequently, no prior knowledge about the underlying equations is needed or utilized in the proposed method. With the assistance of the deep-learning based PDE discovery method, on the one hand, we are able to validate the method by quantitative comparison of discovered PDEs and theoretical PDEs for long-term behaviors of viscous gravity currents both in data space and scientific semantic space. On the other hand, it is also possible for us to transform raw simulation data for describing short-term behaviors from data space into discovered PDEs in scientific semantic space. Subsequently, by quantitatively comparing discovered PDEs with theoretical PDEs, the hidden mechanisms can be elucidated, which can be expressed as compensation terms for modifying the original theoretically derived PDEs to capture short-term behaviors in viscous gravity currents.

The remainder of this paper is organized as follows. In section 2, the two-dimensional microscopic governing equation and numerical solution of viscous gravity current related to training data preparation are introduced. Theoretical derivations of one-dimensional governing equations for viscous gravity currents are also briefly reviewed as references for subsequential quantitative comparisons. Then, the proposed deep-learning based PDE discovery framework is introduced. In section 3, PDE discovery results based on the deep-learning framework for both long-term and short-term regimes of viscous gravity currents are demonstrated, and the effects of data noise are discussed. Finally, the conclusions are drawn in section 4.

## 2 Methodology

### 2.1 Preparation for training data of viscous gravity current

Generally, for investigating fluid mechanics, such as viscous gravity currents, three fundamental approaches usually exist, i.e., theoretical analysis, physical experiments, and numerical simulation[33]. Particularly, theoretical analysis offers strong explainability and scientific consistency. Macroscopic equations can be rigorously derived based on fundamental laws and rational assumptions, and conveniently extended for investigating similar physical processes. The prediction accuracy of theoretical analysis, however, depends greatly on the assumption validity. On the other hand, although physical experiments and microscopic numerical simulations can provide a large amount of observation data and accurate prediction results, they are both time-consuming and possess poor transferring capability. Different from conventional theoretical analysis for obtaining macroscopic equations, we aim to find one-dimensional macroscopic PDEs in scientific semantic space to describe the evolving behaviors of front height based on raw observation data in data space. In this work, the raw observation data are extracted from refined numerical simulation, also referred to as microscopic simulation.

Particularly, we consider a viscous gravity current process in a two-dimensional rectangle vertical fracture with non-penetration boundaries, as shown in Figure 1(a). The permeability and porosity of the fracture are considered as constant, and fluid leak-off, as well as other source terms, are ignored for simplicity. The domain is initially vertically divided by two types of fluid with different density, i.e., $\rho_1$ and $\rho_2$, and different viscosity, i.e., $\mu_1$ and $\mu_2$. Then, determination of how the current front height $h(t,x)$ evolves at macro-scale constitutes the primary aim.

In fact, the fluid motion in this two-dimensional system is governed by Darcy's equation and the continuity equation:

$$\vec{u} = -(k/\mu)(\nabla P - \rho \vec{g}), \tag{1}$$

$$\nabla \cdot \vec{u} = 0, \tag{2}$$

where $k$ is the permeability of fracture; $\rho$ and $\mu$ are the fluid viscosity and density, respectively; $P$ is the pressure; $\vec{g}$ is the gravity acceleration; and $\vec{u}$ is the two-dimensional velocity of fluid.

The interface between two fluids can be tracked with a conservative level-set equation[34]:

$$\frac{\partial C}{\partial t} + \nabla \cdot (C\vec{u}) = 0 \tag{3}$$

$$\frac{\partial C}{\partial \tau} + \nabla \cdot \left[ C(1-C)\frac{\nabla C}{\|\nabla C\|} \right] = \varepsilon \Delta C \tag{4}$$

where $C$ is the level set function or color function; $\vec{u}$ is the two-dimensional velocity of fluid; $\tau$ is the artificial time; and $\varepsilon$ is the artificial viscosity. Note that equation (3) has been written in a conservative form by utilizing the continuity equation (2), and the artificial viscosity $\varepsilon$ is introduced in consideration of numerical stability. The above equations indicate the advection step and the artificial compression step, respectively.

Furthermore, by substituting Darcy's equation (1) into the continuity equation (2), a Poisson equation for pressure independent of fluid velocity can be obtained:

$$\nabla^2 P = \frac{g\mu}{k}\frac{\partial \rho}{\partial z} \tag{5}$$

To obtain the solution of the above two-dimensional system, for each time step, fluid pressure and velocity are first determined by solving the pressure equation (5), and color function $C$ is then updated according to equations (3) and (4). Pressure calculation and evolution of color function are sequentially solved through the simulation timeline.

Numerically, the pressure equation is discretized with the center difference scheme, and the AGMG solver based on the multi-grid technique[35] is adopted to solve the discretized linear system. For the advection step (3), the third-order Runge-Kutta TVD scheme[36] and fifth-order WENO scheme[37,38] are applied for temporal and spatial discretizations, respectively. For the artificial compression step (4), a center difference scheme is adopted, and after several artificial time steps, the color function converges to maintain a constant interface width.

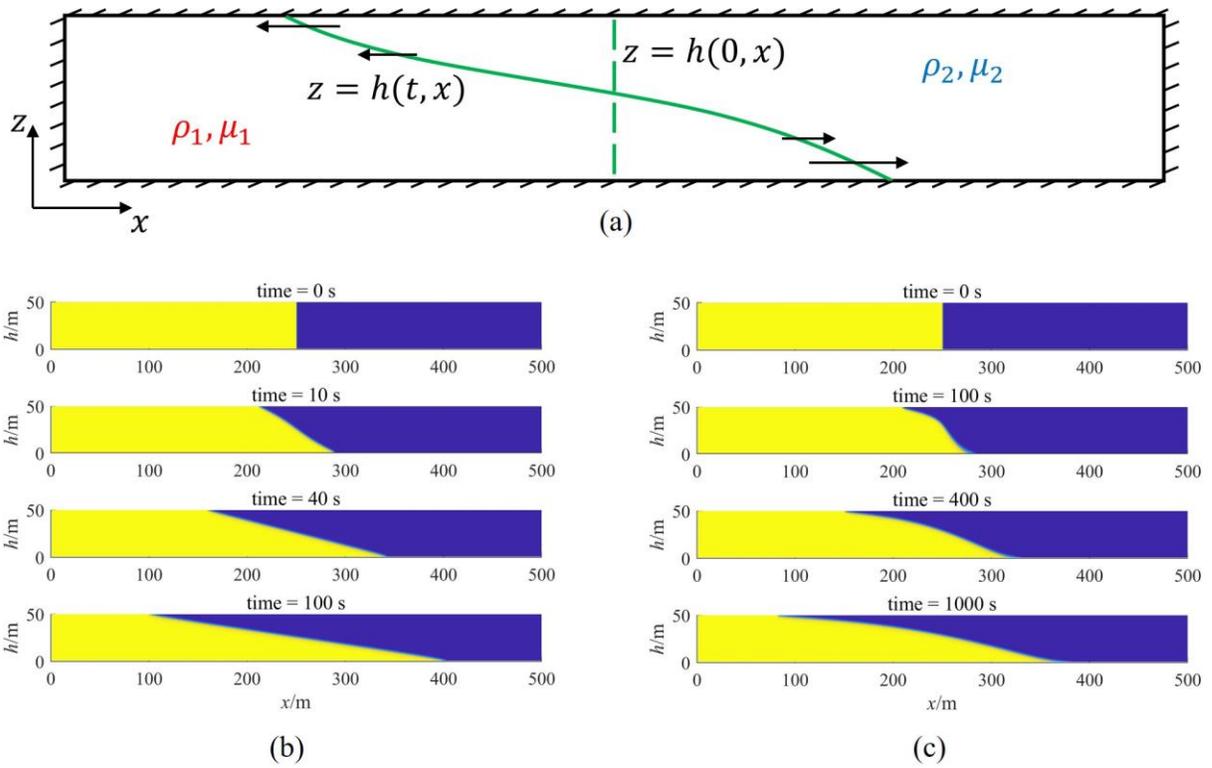

**Figure 1.** Illustrations of viscosity gravity currents. (a) Sketch of viscous gravity current in a rectangle vertical fracture ( $\rho_1 > \rho_2$ ). The green dashed vertical line indicates the initial interface between the two fluids, and the green solid curve $h(t,x)$ indicates the spatio-temporal evolving current front height. (b) Evolution of viscous gravity current for case I, where viscosity of two fluids satisfy $\mu_1 = \mu_2$. The yellow region (level-set function $C=1$) indicates heavy fluid, and the blue region ($C=0$) indicates light fluid. (c) Evolution of viscous gravity current for case II, where viscosity of two fluids satisfy $\mu_1 \gg \mu_2$.

In this work, two typical scenarios of viscous gravity currents are investigated. For case I, viscosity of the heavy fluid is assumed to be equal to that of light fluid, i.e., $\mu_1 = \mu_2$; whereas, for case II, it is assumed that $\mu_1 \gg \mu_2$. Figure 1(b) and 1(c) illustrate the short-term evolving history of current fronts in the two cases. It is observed that the current front tends to be linear-shaped in case I, while a curve-shaped front is seen in case II, during the late time of simulation. By tracking the contour line of level set function $C = 0.5$ in the microscopic simulation results, the spatio-temporal evolving history of current front height $h(t,x)$ is then obtained.

Without loss of generality, related variables, including front height $h$, horizontal location $x$, and time $t$, are normalized through non-dimensional analysis prior to subsequential deep-learning based PDE discovery, and the dimensionless variables are defined as:

$$h^* = \frac{h}{H}; \quad x^* = \frac{x}{H}; \quad t^* = \frac{t\Delta\rho g k}{\mu_1 H} \tag{6}$$

where $H$ is the domain height; and $\Delta\rho = \rho_1 - \rho_2$ is the density difference between the two fluids. It is worth noting that, through the nondimensionalization formula above, changing related physical parameters, including $\Delta\rho$, $g$, $k$, $\mu_1$, and $H$, does not affect the spatio-temporal evolution of dimensionless front height $h^*(t^*, x^*)$. In other words, the obtained PDEs by the deep-learning framework are indeed universal forms for similar processes if simulation data are nondimensionalized by equation (6).

**2.2 Theoretically derived PDEs for viscous gravity currents**

Although the theoretically derived PDEs are obtained based on various assumptions and their applications are usually limited, they can still be regarded as baselines, or asymptotic solutions for the to-be-investigated physical process. In this part, we briefly review the details of theoretically derived PDEs for viscous gravity currents following Gardner et al.'s work[19], which describes the long-term asymptotic behavior of current front height.

Basically, it is assumed that the vertical flow is sufficiently smaller than the horizontal flow, i.e., $u_z \ll k\rho g/\mu$, where $u_z$ indicates the vertical component of fluid velocity. Under this condition, the pressure gradient in the vertical direction can be simplified as $\partial P/\partial z = -\rho g$. Let $P_1$ and $P_2$ denote the fluid pressure in fluid 1 and 2 across a vertical line inside the domain, respectively. Then, the pressure can be analytically expressed as:

$$P_2 = \pi + \rho_2 g (H - z), \quad z > h(x) \tag{7}$$

$$P_1 = \pi + \rho_2 g (H - h) + \rho_1 g (h - z), \quad z > h(x) \tag{8}$$

where $h(x)$ indicates the horizontally varying current front height; $H$ is the total domain height; and $\pi$ is the reference pressure in the horizontal direction.

Then, by utilizing Darcy's law and the mass conservation equation along vertical lines, it is not difficult to find the velocity solution for one fluid:

$$u_{x,1} = k \Delta \rho g \frac{\partial h}{\partial x} \bigg/ \left( \mu_1 + \mu_2 \frac{h}{H-h} \right) \tag{9}$$

Note that, for phase 1, we also have the continuity equation:

$$\frac{\partial h}{\partial t} + \frac{\partial (u_{x,1} h)}{\partial x} = 0 \tag{10}$$

By substituting equation (9) into (10), the PDE of current height can finally be determined as:

$$\frac{\partial h}{\partial t} - k \Delta \rho g \frac{\partial}{\partial x} \left( \frac{h(H-h)}{\mu_1 (H-h) + \mu_2 h} \frac{\partial h}{\partial x} \right) = 0 \tag{11}$$

Considering the two cases in this work, i.e., $\mu_1 = \mu_2$ and $\mu_1 \gg \mu_2$, and nondimensionalizing the above equation based on equation (6), the theoretically derived dimensionless PDEs can be written as:

$$\frac{\partial h^*}{\partial t^*} = \frac{\partial}{\partial x^*} \left( h^* \frac{\partial h^*}{\partial x^*} - h^{*2} \frac{\partial h^*}{\partial x^*} \right) \tag{12}$$

$$\frac{\partial h^*}{\partial t^*} = \frac{\partial}{\partial x^*} \left( h^* \frac{\partial h^*}{\partial x^*} \right) \tag{13}$$

In this work, two regimes are investigated for both cases, i.e., long-term and short-term regimes. As mentioned previously, long-term regimes imply that nondimensional vertical velocity is sufficiently small, i.e., $u_z^* = u_z \mu_1 / (\Delta \rho g k) \ll 1$ [19]. Here, we distinguish these two regimes through another more intuitive indicator, i.e., the dimensionless horizontal spanning distance $\Delta x^* = \|x|_{h^* \to 0} - x|_{h^* \to 1}\|$ of the current front interface. The long-term regime is defined as $\Delta x^* > 6$, and the short-term regime as $0 \leq \Delta x^* \leq 6$. Under this condition, for both cases, the calculated average nondimensional vertical velocities of long-term regimes are less than 0.05, which is acceptably small. Furthermore, for subsequential deep-learning, the amount of training data (spatio-temporal points of front height) of the two regimes is set as the same size.

**2.3 Deep-learning based PDE discovery**

In fact, if ignoring source/sink terms and inhomogeneity of vertical permeability, one-dimensional macroscopic equations for all viscous gravity currents can be expressed in a unified conservative form, i.e., $\partial h^* / \partial t^* + \partial F^* / \partial x^* = 0$, where $F^*$ indicates dimensionless flux. Then, for a to-be-investigated viscous gravity current process, the goal of discovering a concise PDE for front height evolution can be converted into a sparse regression problem of approximating flux $F^*$ through basic algebraic combinations of height $h^*$ and its arbitrary-order derivatives, such as $\partial h^* / \partial x^*$ and $\partial^2 h^* / \partial x^{*2}$. Above all, prior to regression, flux and various derivatives should be calculated in advance based on raw training data. Particularly, flux $F^*$ can be calculated through integration $F^*(t^*, x^*) = \int_{x_0^*}^{x^*} \left(-\partial h^* / \partial t^*\right) dx^*$.

Based on this insight, a natural approach is to calculate the derivatives and integration. First, one may fit the data with smoothing techniques, such as Gaussian process, and then the derivatives and integration can be calculated using conventional numerical difference and integration schemes. However, it has been proven in previous literature[39] that these conventional approaches are sensitive to data noise. Usually, there is no guarantee about the quality of collected data in reality. Therefore, when investigating a physical process without any prior knowledge, the PDE discovery approach should be primarily expected to be robust enough while accuracy of data reconstruction, as well as derivative and integral calculation, can also be maintained. To solve the mentioned issues, here we apply a well-balanced framework based on deep-learning for discovering macroscopic PDEs of viscous gravity currents, as illustrated in Figure 2. Basically, it consists of the following three main steps:

(1) Data reconstruction or surrogate training. As previously mentioned, in practical scenarios, raw training data can be noisy, or deviate from ground truth, and randomly distributed in the spatio-temporal domain, which will influence derivative/integral calculations and PDE discovery. Therefore, a deep neural network $h^* = NN(t^*, x^*; \theta)$, where $\theta$ is the training parameters in the DNN, is first trained by available data to reconstruct the spatio-temporal evolution of front height. Note that, in the DNN, the nondimensional time and location are the only two input variables in the first layer (or input layer), and the nondimensional height is the only output variable in the last layer (or output layer). Theoretically, the neural network can fit any complex mapping relationship. Compared to other reconstruction methods, neural-network based data reconstruction shows good anti-noise and global fitness capability[10]. In addition, the neural network can effectively generate large amounts of meta-data, which is essential for subsequent integral calculation[18].

(2) Semantic fragments construction or derivative/integral calculation. As the neural-network based surrogate is built up, derivatives of various orders are calculated by automatic differentiation if a smooth activation function is adopted for neural network training. Moreover, sufficient meta-data are generated on a regular lattice, and numerical integration is accomplished based on conventional approaches, such as Gauss-Legendre quadrature. Within this step, flux $F^*$, original variable $h^*$ and its various spatial derivatives, such as $\partial h^*/\partial x^*$, $\partial^2 h^*/\partial x^{*2}$ and $\partial^3 h^*/\partial x^{*3}$, can be obtained simultaneously, and these variables are considered as semantic fragments ready for the subsequent step.

(3) Semantic integration or sparse regression. Within this step, the possible semantic fragments are discovered and form the eventual explainable governing equation. Generally, regression methods with parsimony constraints, such as LASSO and genetic algorithm, can be employed to discover the possible PDE terms and corresponding coefficients. Particularly, the genetic algorithm (GA)[7] is adopted in this work. Compared to other sparse regression methods[9,14,40] that must specify a large number of potential terms and their combinations in the candidate library, the main advantage of GA is that the scale of the candidate library can be much smaller, which is composed of several basic genes, and numerous combinations can be achieved by automatic generation evolution. For example, a high-order derivative combination of $(h^*)^3 (\partial h^*/\partial x^*)^2$ can be recursively generated through mutation and cross-over operations during the GA process with only two basis genes $h^*$ and $\partial h^*/\partial x^*$. As a consequence, the GA-based deep-learning framework does not need to include the true terms of the underlying equation in the initial candidate library, which greatly increases computation efficiency and sparse regression accuracy.

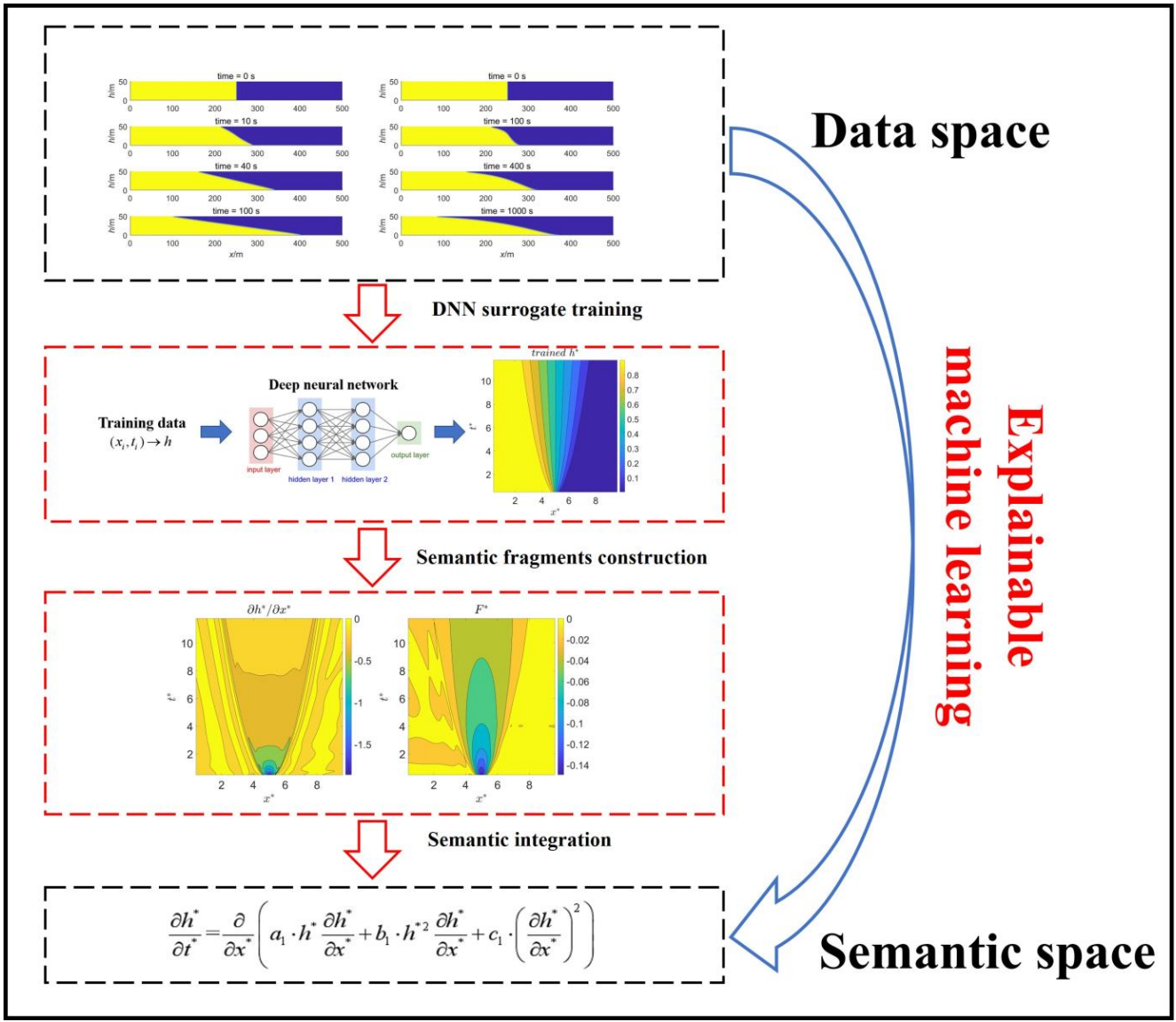

**Figure 2.** Flowchart of the deep-learning based PDE discovery framework. In order to transform raw simulation data in data space into explainable physical laws (PDEs) in semantic space, the following three steps are necessary. (1) Data reconstruction: training the surrogate based on the deep neural network (DNN); (2) constructing semantic fragments: calculating spatial derivatives and integrated flux; and (3) semantic integration: finding a parsimonious PDE based on sparse regression methods.

# 3 Results and Discussion

## 3.1 Validation for long-term regimes

First, to validate the proposed deep-learning framework for discovering macroscopic PDEs, microscopic simulation data of long-term regimes are utilized for investigation. Training details and parameter settings used in this work during the three steps are listed as follows:

(1) For the data reconstruction step, a five-layer fully-connected neural network with 100 neurons per layer, and soft-plus activation function, are used for training. For each case, a total of 80% of $500 \times 500$ spatial-temporal points are chosen as training data. After training approximately 100,000 epochs, the mean square error (MSE) between raw data and prediction results decreases to a minimal level of $10^{-7}$, which is considered to be sufficiently converged.

(2) For derivative/integration calculation, $300 \times 300$ lattice points of the inner spatial-temporal domain are selected, and a five-point Gauss-Legendre quadrature scheme is adopted here to calculate flux. In order to improve the regression performance, a large amount of trivial samples out of the mentioned 90,000 lattice points, e.g., $h^* \approx 1$ and $h^* \approx 0$, need to be omitted because flux is also trivial in these regions, i.e., $F^* \approx 0$.

(3) For the sparse regression step, four basic genes are chosen as the candidate library, i.e., $\left(h^*, \partial h^*/\partial x^*, \partial^2 h^*/\partial x^{*2}, \partial^3 h^*/\partial x^{*3}\right)$. Population size is set as 200 per generation, and genes are randomly combined with each other as the initial generation. Mutation rate and cross-over rate are set as 20% and 80%, respectively. Fitness function is defined as $Fitness = -\sum \left|F^*_{ref} - F^*_{ga}\right|/N - \lambda \sum a_i/|a_i|$, where $N$ is the number of sample data; and $a_i$ is the coefficient of the $i$-th derivative combination. The first term indicates data fitness between reference flux value $F^*_{ref}$ and estimated value $F^*_{ga}$, and the last term indicates the parsimony constraint, i.e., the total number of terms, which is controlled by hyper-parameter $\lambda$. In this work, $\lambda$ is set as 0.002 for most cases.

It is worth noting that there exist several hyper-parameters in the deep-learning framework, such as number of hidden layers, choosing strategy of candidate library, and parsimony constraint coefficient. Most of these parameters are determined by experience via an ad hoc procedure. For instance, the number of neuron layers in the neural networks during the data reconstruction step is determined when further adding the number of layers does not significantly affect the eventual data loss. Sensitivity studies with different high-order derivatives are carried out in constituting the candidate library, and it is found that there is no necessity to introduce derivatives over third-order in the candidate library. The parsimony constraint coefficient is so determined that the regression quality is achieved while not sacrificing the parsimony (with as small a number of PDE terms as possible). Above all, the primary objective of training the deep-learning framework is to discover an accurate yet parsimonious PDE, so that the prediction based on the PDE through posterior tests can well match the raw training data and the learned PDE can reveal the physical insight.

Eventually, for describing long-term behaviors, the obtained deep-learning based PDEs (DL-PDEs) of the two cases are expressed as follows:

$$\frac{\partial h^*}{\partial t^*} = \frac{\partial}{\partial x^*}\left(0.872 \cdot h^* \frac{\partial h^*}{\partial x^*} - 0.881 \cdot h^{*2} \frac{\partial h^*}{\partial x^*}\right) \tag{14}$$

$$\frac{\partial h^*}{\partial t^*} = \frac{\partial}{\partial x^*}\left(0.988 \cdot h^* \frac{\partial h^*}{\partial x^*} + 0.0325 \cdot h^{*2}\right) \tag{15}$$

To provide insight into how GA works to find the optimized solution, Table 1 lists the GA solution paths for the two cases. For case I, the discovered best child for flux in the first generation is found to be a set of $\{h^*, h^{*2}, \partial^2 h^*/\partial x^{*2}\}$, which contains second-order derivative $\partial^2 h^*/\partial x^{*2}$. However, after only three generations, the optimized genomes are converged to $\{h^* \partial h^*/\partial x^*, h^{*2} \partial h^*/\partial x^*\}$. Quantitatively, the fitness loss eventually decreases to 0.0048 from 0.0077 of the first generation. From the solution path, it is clearly seen that the initial candidate library, i.e., $\{h^*, \partial h^*/\partial x^*, \partial^2 h^*/\partial x^{*2}, \partial^3 h^*/\partial x^{*3}\}$, does not need to be complete to include eventual terms in this approach. Indeed, to construct a complete candidate library would usually mean to include an overwhelmingly large number of possible terms, which may greatly increase the computational cost and diminish the performance of sparse regression. It is clear that, to describe the long-term behaviors of viscous gravity currents, the form of flux is highly concise. Since in each generation, a total of 200 children are generated for evolution, the optimized solution can be determined without much difficulty.

**Table 1.** GA solution path of long-term regimes for two cases.

| Generations for case I | Structure of discovered best child for case I | Generations for case II | Structure of discovered best child for case II |
|:---:|:---:|:---:|:---:|
| 1 | $h^*, h^{*2}, \dfrac{\partial^2 h^*}{\partial x^{*2}}$ | 1 | $h^*, h^* \dfrac{\partial h^*}{\partial x^*}$ |
| 2 | $h^*, h^{*2}$ | 4 | $h^{*3}, h^* \dfrac{\partial h^*}{\partial x^*}$ |
| 4 | $h^* \dfrac{\partial h^*}{\partial x^*}, h^{*2} \dfrac{\partial h^*}{\partial x^*}$ | 5 | $h^{*2}, h^* \dfrac{\partial h^*}{\partial x^*}$ |
| 100 | $h^* \dfrac{\partial h^*}{\partial x^*}, h^{*2} \dfrac{\partial h^*}{\partial x^*}$ | 100 | $h^{*2}, h^* \dfrac{\partial h^*}{\partial x^*}$ |

It can be seen that the trained long-term DL-PDEs (14) and (15) are highly similar to the theoretically derived equations (12) and (13), while coefficients are slightly dissimilar for both cases, and an additional term appears for case II. Particularly for case I, the corresponding coefficients of dominated terms $h^* \partial h^*/\partial x^*$ and $h^{*2} \partial h^*/\partial x^*$ deviate approximately 12% from the theoretical PDE. For case II, the coefficients of dominated term $h^* \partial h^*/\partial x^*$ deviate approximately only 2% from the theoretical PDE, and the coefficient of the compensation term is 0.0325, which is also much smaller compared to the dominated term.

Because the long-term DL-PDEs (14) and (15) are directly learned from the microscopic simulation data, it is difficult to discern the error sources resulting in the deviations between DL-PDEs and theoretical-PDEs. The deviations may be attributed to the training error of deep-learning or the invalidity of assumptions for deriving theoretical PDEs. To clarify this problem, it is necessary to carry out further quantitative comparisons among microscopic simulation data, DL-PDEs, and theoretical PDEs. However, equations (12)-(15) are all expressed as PDEs in scientific semantic space, while simulation results are expressed in data space. Since it is impossible to directly compare two different spaces, one must solve equations (12)-(15) numerically, and compare the posterior results with microscopic simulation results in data space. Particularly, to obtain a numerical solution for deep-learning based PDEs, the center difference scheme is first applied for calculating the flux in the cell center, and then the WENO scheme is utilized for flux reconstruction similar to solving equation (4).

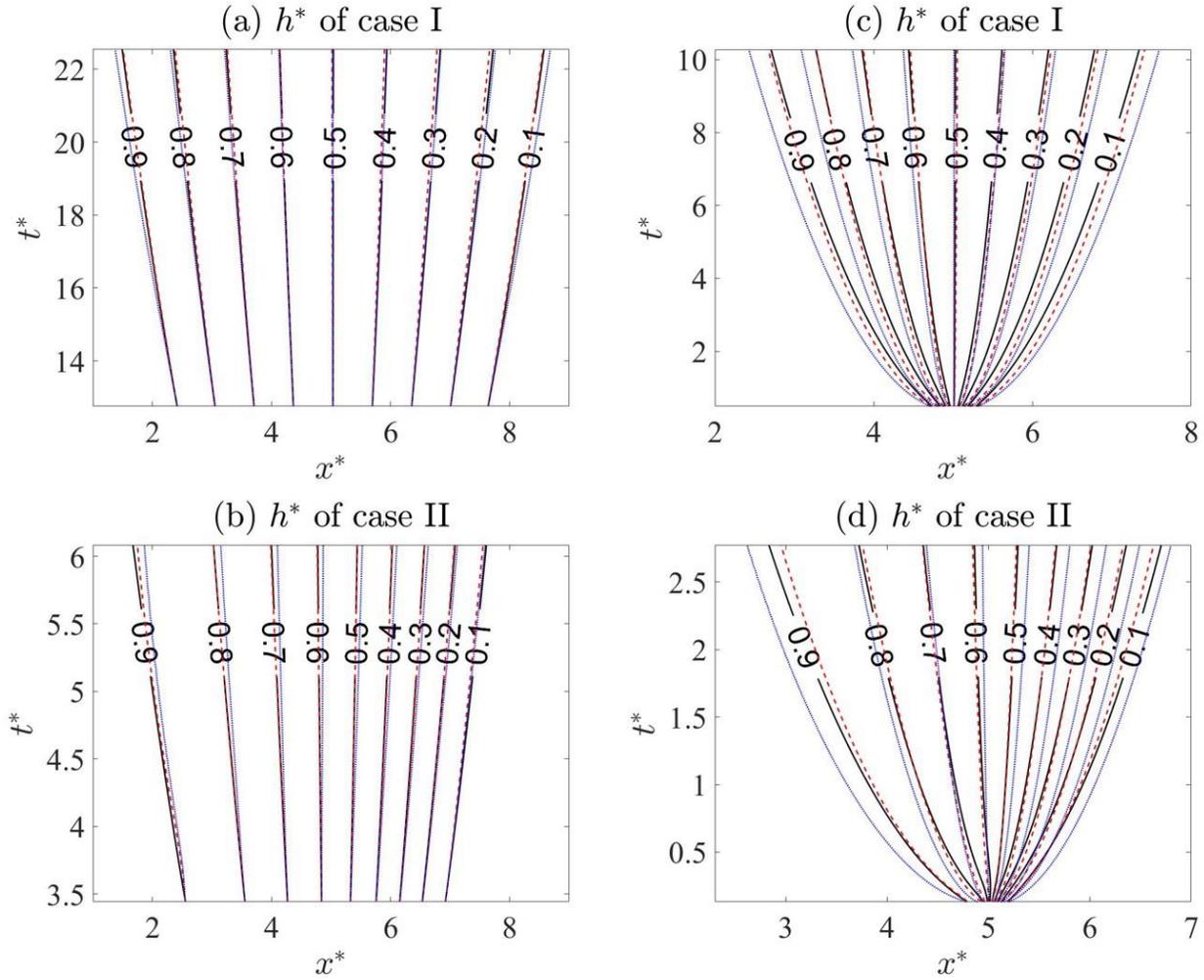

**Figure 3.** Contour lines of estimated evolving current front height $h^*$ in posterior tests for two cases: (a) long-term behaviors of case I; (b) long-term behaviors of case II; (c) short-term behaviors of case I; (d) short-term behaviors of case II. The black solid lines are reference values extracted from two-dimensional microscopic simulation results, the long red dashed lines are posterior results of deep-learning based PDEs, and the short blue dashed lines are posterior results of theoretical PDEs.

The posterior results are illustrated in Figure 3(a) and 3(b). It is clear that the characteristic contour lines of the three data sets are close to each other. Quantitatively, for case I, the total relative errors are estimated as 0.45% and 0.75% for DL-PDE and theoretical PDE, respectively. For case II, the errors are 0.47% and 1.1%, respectively. Therefore, the prediction results of DL-PDEs achieve a superior match with the microscopic simulation results compared to those of theoretical PDEs, irrespective of the training error. This finding implies that modifications are necessary, including smaller coefficients and secondary terms, to better describe viscous gravity currents, even for long-term regimes. More importantly, it is proven that the proposed deep-learning based framework is highly capable and accurate to discover macroscopic equations from raw microscopic simulation data.

**3.2 PDE discovery for short-term regimes**

In this part, short-term behaviors of viscous gravity currents are investigated. We aim to discover the macroscopic PDEs for describing short-term behaviors, and elucidate the hidden macroscopic mechanisms to identify how deviation between theoretical assumptions and actual situations influences the macroscopic equations. Here, the training parameters of deep-learning are similar to those for long-term DL-PDEs. The eventual learned PDEs for cases I and II are given as follows:

$$\frac{\partial h^*}{\partial t^*} = \frac{\partial}{\partial x^*}\left(a_1 \cdot h^* \frac{\partial h^*}{\partial x^*} + b_1 \cdot h^{*2} \frac{\partial h^*}{\partial x^*} + c_1 \cdot \left(\frac{\partial h^*}{\partial x^*}\right)^2\right) \qquad (16)$$

$$\frac{\partial h^*}{\partial t^*} = \frac{\partial}{\partial x^*}\left(a_2 \cdot h^* \frac{\partial h^*}{\partial x^*} + b_2 \cdot h^* \left(\frac{\partial h^*}{\partial x^*}\right)^2 + c_2 \cdot h^{*4} \frac{\partial h^*}{\partial x^*}\right) \qquad (17)$$

where $a_1 = 0.894$; $b_1 = -0.881$; $c_1 = 0.0757$; and $a_2 = 1.01$; $b_2 = 0.215$; $c_2 = -0.571$.

Compared to the DL-PDEs of long-term regimes, it is evident that flux expressions of equations (16) and (17) are more complex, and the GA solution path indicates that many more generations (57 for case I, and 77 for case II) are necessary for convergence, as demonstrated in Table 2. Indeed, at the early stage of viscous gravity currents, the fluid field has not yet been fully developed, and theoretical assumptions are too ideal to reflect the actual scenario, suggesting that a more complex high-dimensional nonlinear mapping relationship is required for describing short-term spatio-temporal evolution of current front height.

**Table 2.** GA solution paths of short-term regimes for two cases.

| Generations for case I | Structure of discovered best child for case I | Generations for case II | Structure of discovered best child for case II |
|---|---|---|---|
| 1 | $\dfrac{\partial h^*}{\partial x^*}$ | 1 | $h^*, h^* \dfrac{\partial^2 h^*}{\partial x^{*2}} \dfrac{\partial^3 h^*}{\partial x^{*3}}, h^{*2}, h^* \dfrac{\partial h^*}{\partial x^*}, \left(\dfrac{\partial h^*}{\partial x^*}\right)^2$ |
| 2 | $h^*, \dfrac{\partial h^*}{\partial x^*}, h^{*2}$ | 3 | $\dfrac{\partial^3 h^*}{\partial x^{*3}}, h^*, h^* \dfrac{\partial h^*}{\partial x^*}, h^{*3}$ |
| 3 | $h^{*2}\left(\dfrac{\partial h^*}{\partial x^*}\right)^3, h^* \dfrac{\partial h^*}{\partial x^*}, h^{*2} \dfrac{\partial h^*}{\partial x^*}$ | 4 | $h^*, h^* \dfrac{\partial h^*}{\partial x^*}, h^{*2}$ |
| 20 | $h^* \dfrac{\partial h^*}{\partial x^*}, h^{*2} \dfrac{\partial h^*}{\partial x^*}$ | 6 | $h^*, h^* \dfrac{\partial h^*}{\partial x^*}, h^{*2}, h^{*2}\left(\dfrac{\partial h^*}{\partial x^*}\right)^2$ |
| 57 | $h^* \dfrac{\partial h^*}{\partial x^*}, h^{*2} \dfrac{\partial h^*}{\partial x^*}, \left(\dfrac{\partial h^*}{\partial x^*}\right)^2$ | 18 | $h^{*3} \dfrac{\partial h^*}{\partial x^*}, h^* \dfrac{\partial h^*}{\partial x^*}, h^*\left(\dfrac{\partial h^*}{\partial x^*}\right)^2$ |
| 100 | $h^* \dfrac{\partial h^*}{\partial x^*}, h^{*2} \dfrac{\partial h^*}{\partial x^*}, \left(\dfrac{\partial h^*}{\partial x^*}\right)^2$ | 77~100 | $h^* \dfrac{\partial h^*}{\partial x^*}, h^*\left(\dfrac{\partial h^*}{\partial x^*}\right)^2, h^{*4} \dfrac{\partial h^*}{\partial x^*}$ |

By comparing the above deep-learning trained PDEs (DL-PDE) with theoretically derived equations (12) and (13), the following findings are determined:

(1) For both cases, it is clear from equations (16) and (17) that high-order derivatives, i.e., $\partial^2 h^*/\partial x^{*2}$ and $\partial^3 h^*/\partial x^{*3}$, make no contribution to the flux construction. It seems that flux can be considered as a bivariate function of $h$ and $\partial h^*/\partial x^*$. As mentioned previously, high-order derivatives temporally appear in the GA solution path. However, through fitness selection, these terms are automatically eliminated from the optimized genomes. This fact is also verified from DL-PDEs (14) and (15) of the long-term regime, as well as theoretical PDEs (12) and (13). In fact, in viscous gravity current processes, these two terms have strong physical meanings according to theoretical analysis, as illustrated in section 2.2. Specifically, $h$ represents vertical length allowing flux transfer, and $\partial h^*/\partial x^*$ is proportional to average horizontal velocity.

(2) Flux terms in theoretical PDEs are exactly captured in DL-PDEs, regardless of slightly different coefficients. Particularly for case I, coefficients of $h^*\partial h^*/\partial x^*$ and $h^{*2}\partial h^*/\partial x^*$ are 0.894 and -0.881, while being 1 and -1 in theoretical PDEs, respectively. Obviously, the characteristic of mutual additive inverse for these two coefficients is not an accidental consequence, but instead reflects the symmetry characteristic of the solution, i.e., $F\left(h^*, \partial h^*/\partial x^*\right) = F\left(1-h^*, \partial h^*/\partial x^*\right)$. It is worth noting that this relationship is not satisfied in case II. For case II, the coefficient of leading term $h^*\partial h^*/\partial x^*$ is 1.01, which is very close to the theoretical value of 1.

(3) Excluding the different coefficients of leading terms, it is observed that DL-PDEs provide additional terms for two cases, i.e., $0.0757 \cdot \left(\partial h^*/\partial x^*\right)^2$ for case I and $0.215 \cdot h^* \left(\partial h^*/\partial x^*\right)^2 - 0.571 \cdot h^{*4} \partial h^*/\partial x^*$ for case II. It should be emphasized that theoretical PDEs are derived for describing long-term behaviors of viscous gravity currents; whereas, in this work, we extend the training data to the whole range, including short-term behaviors. Therefore, we aim to find a global solution for viscous gravity currents, which implies that the additional terms in DL-PDEs and modified coefficients of leading terms may be considered as compensation terms for describing short-term range behaviors.

For quantitative comparisons, posterior tests are also carried out, and the evolving front height of two-dimensional simulation results, DL-PDEs, and theoretical PDEs are illustrated in Figure 3(c) and 3(d). It is worth noting that it is not guaranteed that the deep-learning based PDEs are exactly numerically stable because of training error, and the negative diffusion term in equations (16) and (17) at the early time can dominate the process at the early time, where $-\partial h^*/\partial x^*$ is very large and leads to computation collapse, which is a consequence of the training error tolerance. To solve this issue, necessary artificial viscosity is introduced to counter the negative diffusion under these circumstances. To illustrate this, let us take equation (16) as an example. Equation (16) can be rewritten as follows:

$$\frac{\partial h^*}{\partial t^*} = \left(a_1 + 2b_1 h^*\right)\frac{\partial h^*}{\partial x^*}\frac{\partial h^*}{\partial x^*} + \left(a_1 h^* + b_1 h^{*2} + 2c_1 \frac{\partial h^*}{\partial x^*}\right)\frac{\partial^2 h^*}{\partial x^{*2}} \qquad (18)$$

The two terms on the right-hand side can be regarded as parametric convection and diffusion terms, respectively. In this case, if the coefficient of the diffusion term is positive, i.e., $a_1 h^* + b_1 h^{*2} + 2c_1 \partial h^*/\partial x^* > 0$, then the system is numerically stable. If the coefficient is negative, however, the system will break down due to anti-diffusion. According to our numerical experiments, it is found that the system is more likely to collapse where $-\partial h^*/\partial x^*$ is very large, which corresponds to the early time in the viscous gravity current.

To suppress the non-physical anti-diffusion terms, artificial viscosity $\eta$ is introduced for deep-learning based PDEs when needed, which is written as:

$$\frac{\partial h^*}{\partial t^*} - \frac{\partial}{\partial x^*}\left(a_1 \cdot h^* \frac{\partial h^*}{\partial x^*} + b_1 \cdot h^{*2} \frac{\partial h^*}{\partial x^*} + c_1 \cdot \left(\frac{\partial h^*}{\partial x^*}\right)^2\right) = \eta \frac{\partial^2 h^*}{\partial x^{*2}} \qquad (19)$$

where $\eta = -\min\left(0, a_1 h^* + b_1 h^{*2} + 2c_1 \partial h^*/\partial x^*\right)$. Similar remedies are adopted for other DL-PDEs when needed during posterior tests, such as in equation (17).

From the characteristic lines of current height in the $t-x$ domain, it is inferred that wave structures of current height in the short-term can be considered as rarefaction waves with time-dependent characteristic velocities. The comparison results demonstrate that prediction results of DL-PDEs match the microscopic simulation results better than those of theoretical PDEs. Quantitatively, for case I, the total relative errors are estimated as 1.06% and 2.75% for DL-PDE and theoretical PDE, respectively. For case II, however, the errors are 0.85% and 2.72%, respectively.

As previously mentioned, deviation between reference data and theoretical results is mainly attributed to the assumptions when deriving theoretical PDEs. In theoretical analysis, it is assumed that vertical velocity is zero for the whole domain, while microscopic simulation results infer that this condition is not fulfilled, particularly near the interface and during the short-term period. It is seen from Figure 4 that, over the whole simulation period, the absolute relative magnitude of vertical velocity ranges markedly from 0.5 to 1. Moreover, the vertical velocity is fairly comparable to the horizontal velocity, which is particularly significant near two interface tails. Therefore, the assumption of zero vertical velocity is not valid, and theoretical PDEs cannot precisely reproduce the evolving process of viscous gravity current. As a consequence, the transverse spreading length in the $t-x$ domain predicted by theoretical PDEs is larger than ground truth. Modified coefficients and compensation terms are necessary to be introduced for improvement. On the other hand, because DL-PDEs are discovered directly from observation data, once training is converged, the posterior performance is reasonably better than that of theoretical PDEs.

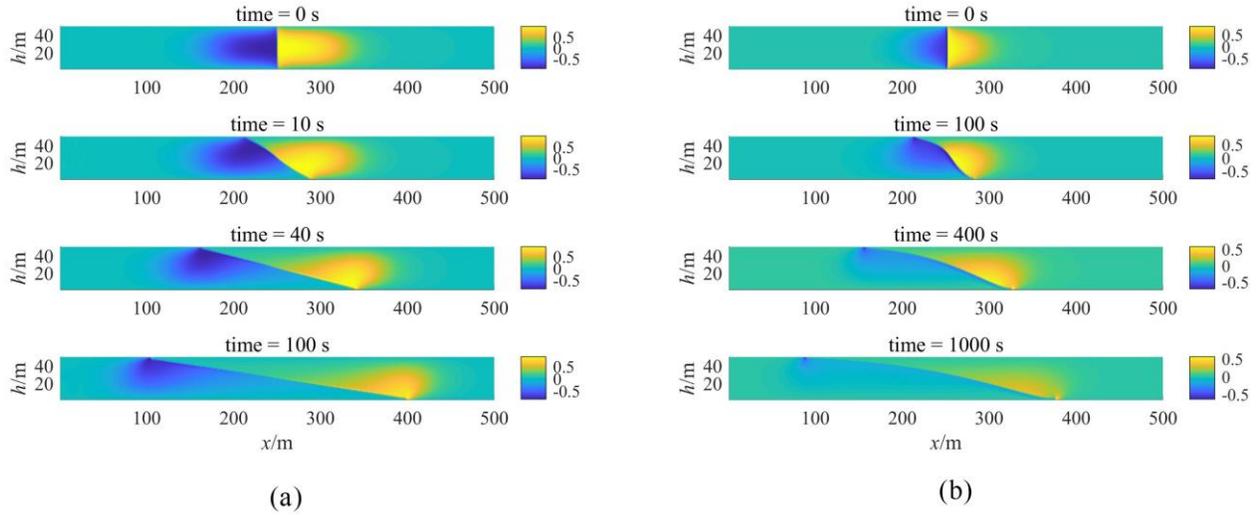

**Figure 4.** Contour of relative magnitude of vertical velocity component $u_z / \sqrt{u_x^2 + u_z^2}$ at various time instances for two cases: (a) case I; (b) case II.

From the above discussions, the following key points can be inferred for the mechanisms hidden in DL-PDEs:

(1) Smaller coefficients in leading terms for case I weaken the strength of the rarefaction wave and adjust theoretical PDEs towards ground truth.

(2) Compensation terms for cases I and II essentially represent negative-diffusion mechanisms, which also suppress the over-spreading trend for theoretical PDEs. For cases I and II, the compensation diffusion terms are $0.151 \cdot \partial h^* / \partial x^* \left( \partial^2 h^* / \partial x^{*2} \right)$ and $0.43 \cdot h^* \, \partial h^* / \partial x^* \left( \partial^2 h^* / \partial x^{*2} \right) - 0.571 \cdot h^{*4} \, \partial^2 h^* / \partial x^{*2}$, respectively.

Furthermore, since in this work we have $0 \leq h^* \leq 1$ and $\partial h^*/\partial x^* \leq 0$, the coefficients of the compensation diffusion terms are negative for both cases.

## 3.3 Effects of data noise

Usually, data noise is a nonnegligible factor when collecting raw training data, especially in physical experiments. Excessive data noise can dramatically affect the validity of the discovered PDEs. To elucidate the robustness of the proposed deep-learning framework, in this part raw simulation data with appended random noise are utilized to discover the PDEs for short-term regimes. Here, case II is taken as an example, and Figure 5(a) and 5(c) illustrate the 3D surfaces of raw training data with 10% and 20% non-biased noise. Although a large number of burrs due to data noise are observed, the basic evolving trend of dimensionless current front height can be still maintained in the raw data to some extent. Figure 5(b) and 5(d) illustrate the DNN surrogate constructed from simulation data with 10% and 20% non-biased noise, respectively. In this part, the training parameters of DNN are all the same as those mentioned in section 3.1. Due to the data noise, the training loss finally keeps a level of $10^{-4}$ after training 100,000 epochs. Clearly, burrs due to data noise no longer exist in the surrogate, which proves that DNN is actually a powerful global smoother to construct a smooth-enough surrogate for subsequential derivative/integral calculations.

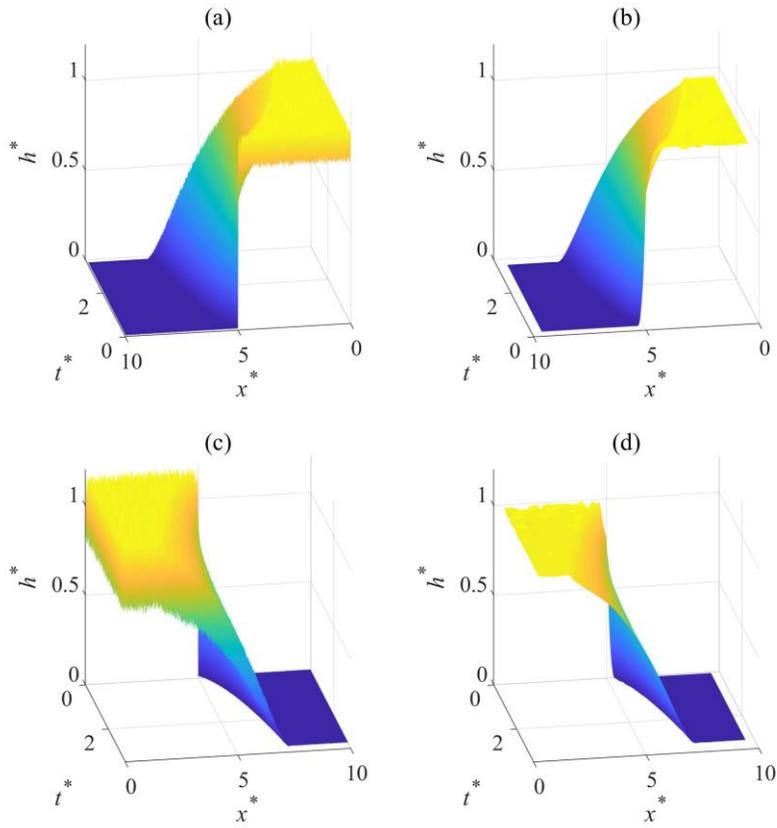

**Figure 5.** 3D illustrations of raw training data ((a) and (c)) and DNN surrogate ((b) and (d)). (a) and (b): 10% data noise; (c) and (d): 20% data noise.

Table 3 lists the discovered PDEs based on the deep-learning framework for these cases. Essentially, as the magnitude of appended noise increases, the fitness loss in the genetic algorithm monotonically increases. Even so, for case I with 10% data noise, the form of discovered PDE is exactly the same as that discovered from clean simulation data. The relative error of the coefficients of leading terms for the clean case and the 10% noise case is smaller than 2%, and the relative error of compensation terms is approximately 15%. On the other hand, for case I with 20% noise, it seems that the proposed deep-learning framework fails to find the aforementioned compensation terms. Instead, the coefficients of leading terms are much smaller than those of the clean data case. As mentioned previously, smaller coefficients and compensation terms both contributed to weaker rarefaction mechanisms. Therefore, even with 20% data noise, the discovered PDE also demonstrates the necessity of leading terms and modifications in the effective direction.

For case II, it is clear that despite data noise, all of the discovered PDEs possess consistent forms. The coefficient of the leading term is always approximately 1, and the relative errors of compensation terms are less than 20%. It is also clear that, although the training loss in this part is much higher than the results with no data noise, the basic structure of the obtained DL-PDEs can still be maintained. Above all, through the numerical tests in this section, it is proven that the proposed DLGA framework is quite robust to discover the explainable PDEs even within 20% data noise for the viscous gravity current processes. From the authors' point of view, the robustness of the proposed framework is to a great extent attributable to the strong data fitting and smoothing capability of the neural network. It should be noted that although the data noise is assumed to be non-biased in this part, advanced approaches, such as the 'robust DLGA' framework[41], can be utilized to solve more complex problems that contain biased data noise.

**Table 3**. DL-PDEs for viscous gravity current (short-term regimes) at different noise levels.

| Noise level | Case I | Fitness loss | Case II | Fitness loss |
|---|---|---|---|---|
| **0%** | $\frac{\partial h^*}{\partial t^*} = \frac{\partial}{\partial x^*}\left(0.894 h^* \frac{\partial h^*}{\partial x^*} - 0.881 h^{*2} \frac{\partial h^*}{\partial x^*} + 0.0757\left(\frac{\partial h^*}{\partial x^*}\right)^2\right)$ | 0.00716 | $\frac{\partial h^*}{\partial t^*} = \frac{\partial}{\partial x^*}\left(1.006 h^* \frac{\partial h^*}{\partial x^*} + 0.215 h^*\left(\frac{\partial h^*}{\partial x^*}\right)^2 - 0.571 h^{*4} \frac{\partial h^*}{\partial x^*}\right)$ | 0.0133 |
| **10%** | $\frac{\partial h^*}{\partial t^*} = \frac{\partial}{\partial x^*}\left(0.912 h^* \frac{\partial h^*}{\partial x^*} - 0.900 h^{*2} \frac{\partial h^*}{\partial x^*} + 0.0876\left(\frac{\partial h^*}{\partial x^*}\right)^2\right)$ | 0.00883 | $\frac{\partial h^*}{\partial t^*} = \frac{\partial}{\partial x^*}\left(1.031 h^* \frac{\partial h^*}{\partial x^*} + 0.245 h^*\left(\frac{\partial h^*}{\partial x^*}\right)^2 - 0.600 h^{*4} \frac{\partial h^*}{\partial x^*}\right)$ | 0.0162 |
| **20%** | $\frac{\partial h^*}{\partial t^*} = \frac{\partial}{\partial x^*}\left(0.737 h^* \frac{\partial h^*}{\partial x^*} - 0.741 h^{*2} \frac{\partial h^*}{\partial x^*}\right)$ | 0.0159 | $\frac{\partial h^*}{\partial t^*} = \frac{\partial}{\partial x^*}\left(1.046 h^* \frac{\partial h^*}{\partial x^*} + 0.233 h^*\left(\frac{\partial h^*}{\partial x^*}\right)^2 - 0.600 h^{*4} \frac{\partial h^*}{\partial x^*}\right)$ | 0.0550 |

## 4 Conclusion

In summary, we proposed a deep-learning framework to elucidate the macroscopic equations of viscous gravity currents based on microscopic simulation data. The problem of discovering macroscopic PDEs is converted to a sparse regression problem of representing flux with primitive variables and corresponding spatial derivatives. With the assistance of the deep neural network and genetic algorithm, PDEs with parsimonious forms for both long-term and short-term behaviors for two typical viscous gravity current processes are obtained. By quantitative comparison between theoretical PDEs and DL-PDEs for describing long-term behaviors, it is proven that the proposed framework is highly capable and accurate for discovering physical laws from raw data. The short-term simulation data are then utilized for discovering unknown macroscopic equations.

We show that the proposed deep-learning based PDE discovery framework is essentially an explainable machine-learning approach. Microscopic simulation results in data space are eventually transformed into parsimonious PDEs in scientific semantic space through the PDE discovery method. With this framework, explainable compensation terms can be introduced to capture the short-term behaviors more accurately. It is found that, in posterior tests, DL-PDEs perform better than the theoretically derived equations. Moreover, the framework is proven to be very robust against approximately 20% non-biased data noise. Consequently, the proposed method is demonstrated to be highly beneficial to compensate for the drawbacks of theoretical assumptions and construct more accurate governing equations for practical physical problems.

Overall, for a certain physical process, obtained experimental or simulation data contain all kinds of information in data space. However, one needs to extract the most valuable and significant content, and ideally determine intrinsic laws in scientific semantic space. This work demonstrates that the proposed deep-learning based PDE discovery framework constitutes a feasible and efficient approach to solve this problem. It is also worth noting that the proposed framework follows a pure data-driven approach, and the hyper-parameters are still manually tuned to achieve parsimony principles, which still need further improvements to consider more prior physics and avoid subjective decisions. Physics-informed structures and Lagrangian dual optimization approaches may be good solutions for these problems. In future research, we will continue to consider these issues and extend the framework for learning multi-dimensional equations and for more challenging scenarios.

## Code availability

The codes used to generate and analyze data are available through Zenodo (https://doi.org/10.5281/zenodo.4587614).